\documentclass[fleqn,twoside]{article}
\usepackage[headings]{espcrc2}

\usepackage{amsmath,amssymb}
\usepackage{dcolumn,array}

\usepackage{epsfig}

\begin{document}

\title{Realistic nuclear Hamiltonian: {\itshape Ab exitu} approach}
\author{A. M. Shirokov\address{Skobeltsyn Institute of Nuclear
    Physics, Moscow State University, Moscow, 119992,
    Russia}$^,$\address[ISU]{Department of Physics and Astronomy, 
Iowa State University, Ames, Ia 50011-3160, USA},
J. P. Vary\addressmark[ISU]$^,$%
\address{Lawrence Livermore National Laboratory, L-414, 7000 East Avenue, 
Livermore, California, 94551, USA}$^,$\address{Stanford Linear Accelerator
  Center, MS81, 2575 Sand Hill 
  Road, Menlo Park, California, 94025, USA},
A. I. Mazur\address{Pacific National University, 
Tikhookeanskaya 136, Khabarovsk 680035, Russia},
T. A. Weber\addressmark[ISU]}
   \begin{abstract}
   Fully-microscopic No-core Shell Model (NCSM) calculations of all stable
   $s$ and
   $p$ shell nuclei are used to determine a realistic
   $NN$ interaction, JISP16,
   describing not only the two-nucleon data but the binding energies and spectra
   of nuclei with $A\leq 16$ as well. The JISP16 interaction, providing
   rapid convergence of the NCSM calculations, is obtained in
   an {\em ab exitu} approach
   by phase-equivalent
   transformations
   of the JISP6 $NN$ interaction.\\


\end{abstract}

\maketitle

To complement the successful but computationally intensive
`{\em ab initio}' No-core Shell Model (NCSM) \cite{Vary3}, we introduce the
`{\em ab exitu}' NCSM.
While the
former has proven very successful for light nuclei when one includes
three-body ($NNN$) forces \cite{NaO,Noggaetal},
the computational complexity motivates us to introduce an approach
that simultaneously minimizes
$NNN$ forces 
while providing
more rapid convergence with a pure nucleon-nucleon ($NN$) force.  We invoke
directly an end-goal of nuclear theory (hence
the term `{\em ab exitu}'), a successful description of nuclear 
properties,
including the available $NN$ data, to develop a new class of $NN$ potentials
that provide accurate descriptions of a broad range of  nuclear data.

To achieve this, we form a union of two recent techniques~--- the $J$-matrix
inverse scattering \cite{Zaitsev,ISTP,JISP6} and the NCSM \cite{Vary3}.
A major ingredient of our approach is the form of the $NN$ interaction (a
small matrix in the oscillator basis), which
is chosen  to provide rapid convergence of many-body observables
within the NCSM.
Indeed, we show below that results up  through $A=16$ obtained
directly with the bare interaction (one that accurately describes the $NN$ data)
are close to those
obtained  with the effective interaction and are very useful to
establish  the confidence region for the binding energy.

Since this is a departure from the more traditional approach, we
motivate our development with
observations concerning the
successful {\em ab initio} approaches to light nuclei.  Indeed
several promising microscopic approaches have been introduced and
tested extensively with realistic $NN$ interactions (see \cite{Bench}
and references therein)
and with realistic $NN + NNN$ interactions  \cite{GFMC,NaO,Noggaetal}.
Progress towards heavier nuclei appears limited only by scientific
manpower and by 
available computers.  However, all
approaches face the exponentially rising computational complexity
inherent in the quantum many-body problem with increasing particle
number and  novel 
schemes are needed to minimize the computational
burden without sacrificing realism and precision.

The earliest and most successful in reaching nuclei beyond $A=4$ is the
Green's-function Monte Carlo (GFMC) approach \cite{GFMC}
whose power  has been used
to determine a sequence of ever-improving $NNN$ interactions 
\cite{GFMC,Argonne-3,Illinois},
in conjunction with highly precise
$NN$ interactions \cite{Argonne}
that fit a wide selection of low-lying
properties of light nuclei up through $A=10$.  In addition, the
Hamiltonians are tested for their predictions in infinite systems 
\cite{Matter}.
According to our usage of terminology here, the application
of GFMC to determine successful $NNN$ interactions is an excellent
example of an {\em ab exitu} approach.

Now, we ask the question whether it is possible to go even further
and search through the residual freedoms of a realistic $NN$
interaction to obtain new $NN$ interactions that satisfy three
criteria: (1) retain excellent descriptions of the $NN$ data; 
(2) provide good fits to light nuclei; 
and (3) provide improved convergence properties within the NCSM.  
The challenge to satisfy this triad of conditions is daunting
and we are able to provide only
an initial demonstration at the present time.

We  are supported by the work  of
Polyzou and Gl\"ockle who demonstrated \cite{Poly} that
a realistic $NN$ interaction 
is equivalent at the $A=3$ level to 
a realistic $NN + NNN$ interaction where the
new $NN$ force is
related to the initial one through a phase-equivalent transformation (PET).
The net
consequence is that properties of nuclei beyond $A=3$ become dependent
on the freedom  within the  transformations at the $A=3$ level.
It seems reasonable then to exploit this freedom and work 
to minimize the need for the explicit introduction 
of three and higher body forces.  However, we do not surmise that
we would be able to eliminate them completely.


We start from the realistic
charge-independent $NN$ interaction JISP6
\cite{JISP6} that provides
an excellent description  of the deuteron properties
\cite{JISP6} and $NN$ scattering data
with $\rm \chi^2/datum = 1.03$ for
the 1992 $np$ data base
(2514 data), and 1.05 
for the 1999 $np$ data base (3058 data)
\cite{chi2-priv}. JISP6 provides
also a very good description of the
spectra of $p$ shell nuclei, but
we find that it overbinds
nuclei with $A\geq 10$.  To eliminate this deficiency,
we exploited PETs to modify
the JISP6 in various partial waves. The resulting interaction,
hereafter referred to as JISP16 since it is fitted in our
{\em ab exitu}
approach  to the spectra and bindings of stable $A\leq 16$ nuclei, can
be obtained from the initial ISTP interaction in the same manner as
JISP6 in Ref. \cite{JISP6} but with a
different set of PET angles. These
angles associated with unitary transformations (see Refs.
\cite{ISTP,JISP6} for details) 
mixing the lowest $s$ and $d$ oscillator
basis states in the coupled $sd$ waves and the   lowest oscillator basis
states in the $^3p_2$, $^3p_1$, $^3p_0$, $^3d_2$ and $^1p_1$ waves are
$\vartheta=-11.0^\circ$, $+5^\circ$, $-6^\circ$, $-10^\circ$, $+25^\circ$ and
$-12^\circ$   respectively. The JISP16 matrix elements in the
oscillator basis with $\hbar\omega=40$ MeV that differ
from those of JISP6,
are presented in Tables
\ref{pot1s0}--\ref{potsd1}. The JISP16 predictions for the deuteron
rms radius $r_d = 1.9643$~fm and quadrupole moment $Q=0.288585$~fm$^2$
slightly differ from the JISP6 results
since the JISP16 and JISP6 PET angle
in the $sd$ coupled waves is
slightly different
($\vartheta=-11.0^\circ$ versus $-11.3^\circ$).
In this paper, we  
include all $NN$ partial waves up to $l=4$ 
and include in Tables
\ref{pot1s0}--\ref{potsd1} the added matrix elements.

\extrarowheight=3pt
\begin{table*}
\begingroup
\tabcolsep=2pt
\caption{JISP16 non-zero matrix elements in $\hbar\omega=40$~MeV units
in the uncoupled
$NN$ partial waves
that differ from the respective JISP6 matrix elements and of the
JISP16 matrices in higher partial waves.}
\label{pot1s0}
\begin{tabular}{@{}>{$}c<{$}>{$}r<{$}>{$}r<{$}c@{\hspace{15pt}}>{$}r<{$}>{$}r<{$}>{$}r<{$}c@{}} \hline
n & \multicolumn{1}{c}{$ V_{nn}^l$}  &
\multicolumn{1}{c}{$V_{n,n+1}^l\!=\!V_{n+1,n}^l$}
          &
{$ V_{n,n+2}^l\!=\!V_{n+2,n}^l$} &
n & \multicolumn{1}{c}{$ V_{nn}^l$}  &
\multicolumn{1}{c}{$V_{n,n+1}^l\!=\!V_{n+1,n}^l$}
          & \multicolumn{1}{c}{$ V_{n,n+2}^l\!=\!V_{n+2,n}^l$}
\\\hline
&\multicolumn{3}{l}{$^1p_1$ partial wave} &
&\multicolumn{3}{l}{$^3p_0$ partial wave}\\
0 &0.4864373541       & -0.2359869829  & 0.3117643519 &
0 & 0.1571004930    & -0.1425039101   & 0.2505691390  \\
1 &-0.1487460250   & -0.1438603014   & &
1 &  -0.2172768679   &  -0.0981725471 \\[1ex] 
&\multicolumn{3}{l}{$^1g_4$ partial wave} &
&\multicolumn{3}{l}{$^3g_4$ partial wave}\\
0 & -0.0159359974 &  0.0110169386   & &
0 &  -0.0762338541  & 0.0498484441   \\
1 &  -0.0229351778 &0.0073206473  & &
1 & -0.1107702854  &  0.0371277135 \\
2 &  -0.0056121168 &  & &
2 & -0.0295683403 \\ \hline        
\end{tabular}
\endgroup

\vspace{4ex}

\caption{Same as in Table \ref{pot1s0} but for  the coupled $NN$ waves.}
\label{potsd1}
\begin{tabular}{@{}l@{\hspace{40pt}}r@{}}\hline
\begin{tabular}{@{}>{$}c<{$}>{$}r<{$}>{$}r<{$}>{$}c<{$}@{}} 
\multicolumn{2}{l}{$sd$ coupled waves}\\ \hline
  & \multicolumn{3}{l}{$V^{ss}_{nn'}$ matrix elements} \\ \cline{1-3}
    n &  \multicolumn{1}{c}{$V_{nn}^{ss}$}
        &
      V_{n,n+1}^{ss}\!=\!V^{ss}_{n+1,n}  \\ \cline{1-3}
0 &  -0.5125432769 &  0.2139078754 \\[1ex] 
  & \multicolumn{3}{l}{$V^{dd}_{nn'}$ matrix elements} \\  \cline{1-3} 
   n &\multicolumn{1}{c}{$ V_{nn}^{dd}$}
          & 
     V_{n,n+1}^{dd}\!=\!V^{dd}_{n+1,n} \\ \cline{1-3} 
0 &    0.0551475852  & -0.0952367414 \\[1ex] 
   & \multicolumn{3}{l}{$V^{sd}_{nn'}=V^{ds}_{n'n}$ matrix elements} \\  \hline
    n &V_{n,n-1}^{sd}\!=\!V^{ds}_{n-1,n}
         &\multicolumn{1}{c}{$ V_{nn}^{sd}=V_{nn}^{ds}$}
      & V_{n,n+1}^{sd}\!=\!V^{ds}_{n+1,n} \\ \hline
0 &      & -0.4035852241   &  0.2003382771  \\
1 &  -0.0464306332  & & \\  
& & \\[-1.ex] \hline
\multicolumn{2}{l}{$pf$ coupled waves}\\ \hline
   &\multicolumn{3}{l}{$V^{pp}_{nn'\vphantom{,}}$ matrix elements} \\
   \hline
    n &\multicolumn{1}{c}{$ V_{n n}^{pp}$}
     & 
          V_{n, n+1}^{pp}\!=\!V_{n+1,n}^{pp}
     & 
          V_{n, n+2}^{pp}\!=\!V_{n+2,n}^{pp}\\ \hline
0& -0.1933759934  &  0.1508436490  &  -0.1072949881 \\
1 &   -0.0277262441 &  0.0964883300 & \\[1ex] 
  &  \multicolumn{3}{l}{ $V^{pf}_{nn'\vphantom{,}}$ matrix elements } \\
   \hline
    n &  
              V_{n, n-1}^{pf}\!=\! V_{n-1,n}^{fp}
       & \multicolumn{1}{c}{$ V_{n n}^{pf}=V^{fp}_{nn}$}
       & 
           V_{n, n+1}^{pf}\!=\!V^{fp}_{n+1, n} \\ \hline
0&     & 0.0195093232 &  0.0020663826  \\
1 &  -0.0252003957   & 0.0236188613 
\end{tabular}
    &
\begin{tabular}{@{}>{$}c<{$}>{$}r<{$}>{$}r<{$}@{}} 
\multicolumn{2}{l}{$dg$ coupled waves}\\ \hline
 &\multicolumn{2}{l}{$V^{dd}_{nn'\vphantom{,}}$ matrix elements}\\ \cline{1-3}
    n &\multicolumn{1}{c}{$ V_{n n}^{dd}$}
     & 
          V_{n, n+1}^{dd}\!=\!V_{n+1,n}^{dd}\\  \cline{1-3} 
0 &   -0.0226611102  &  0.0231171026  \\
1 &  -0.0514940563 & 0.0256493733 \\
2 & -0.0329967376 &  0.0061799968 \\
3&  -0.0002368252 & \\[1ex] 
 & \multicolumn{2}{l}{$V^{gg}_{nn'\vphantom{,}}$ matrix elements}\\ \cline{1-3}
    n &\multicolumn{1}{c}{$ V_{n n}^{gg}$}
     & 
          V_{n, n+1}^{gg}\!=\!V_{n+1,n}^{gg}\\  \cline{1-3} 
0 &  0.0435654902  & -0.0276372780 \\
1 & 0.0537629744  &  -0.0140723375 \\
2 & 0.0079901608  &  \\[1ex] 
&\multicolumn{2}{l}{ $V^{dg}_{nn'\vphantom{,}}$ matrix elements }\\ \cline{1-3}
    n &  
              V_{n, n-1}^{dg}\!=\! V_{n-1,n}^{gd}
       & \multicolumn{1}{c}{$ V_{n n}^{dg}=V^{gd}_{nn}$} \\ \cline{1-3}
0&     &  -0.0392683838     \\
1 &   0.0791431969    &  -0.0874578184 \\
2 &  0.0660805779     & -0.0334474774  \\
3 &    0.0029846726  
\end{tabular}\\ \hline
\end{tabular}
\end{table*}

Our fitting procedure is one of 'trial-and-error' where we worked with
only a few partial waves that we deemed important for these nuclei. 
We fit only the
excitation energies of few lowest $^6$Li levels and the $^6$Li and
$^{16}$O binding energies.
To save time, we performed the NCSM calculations in small enough model
spaces (up to $10\hbar\omega$ for 
$^6$Li and up to $4\hbar\omega$ for $^{16}$O). After
obtaining a reasonable description of these observables, we checked that
the binding energies and spectra of all the rest $s$ and $p$ shell
nuclei are well-described in small model spaces. The results presented
below are obtained in the  {\it ab initio} NCSM calculations with the
obtained $NN$ interaction, the {\it ab exitu} JISP16, in larger model spaces. 
This description of the
binding energies is somewhat worse than the one obtained during the
fit in smaller model spaces, however it is still very reasonable. 
In a future effort, we will perform a thorough search through the
space of possible PETs that should further improve 
the description of nuclear properties while retaining the
excellent description of the $NN$ data.

\begin{figure}
\centerline{\epsfig{file=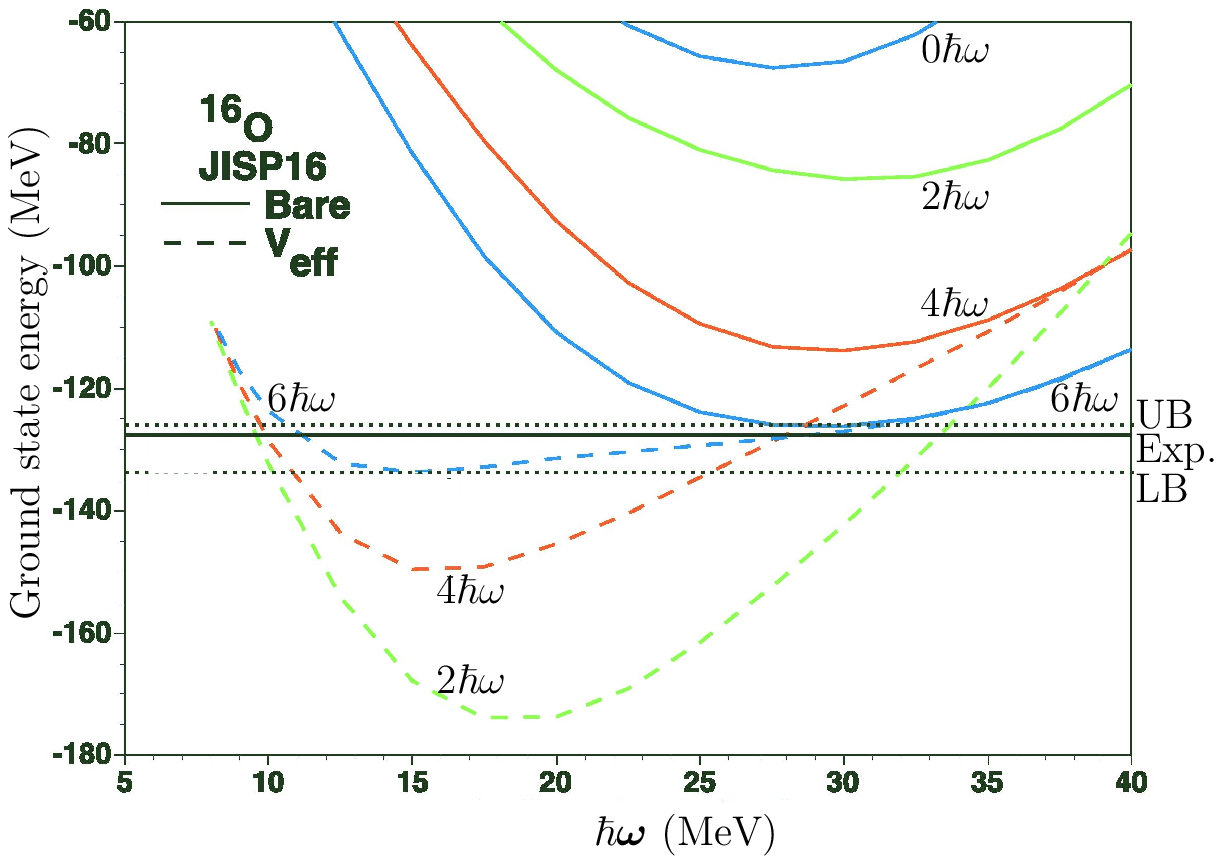,width=.48\textwidth}}
\caption{(Color online) The $\hbar\omega$ dependence of the $^{16}$O ground
state energy obtained with bare JISP16 and effective
interaction based on  JISP16 in a sequence of
$N_{max}\hbar\omega$ model spaces up to $N_{max}=6$; the lines marked
as Exp., UB and LB show the experimental ground state energy, the upper bound
and the suggested lower bound for the NCSM ground state energy predictions.}
\label{hwdep}
\end{figure}

We illustrate our approach with
the $^{16}$O ground
state energy in Fig. \ref{hwdep}. 
The variational principle holds for the bare interaction results; 
hence the upper
bound (UB) for the  ground state energy is the
minimum of its $\hbar\omega$ dependence
in the 
$6\hbar\omega$ model space. In the
calculations with the effective interaction obtained by  the
Lee--Suzuki transformation, the quoted result is  conventionally
associated with the minimum of the
$\hbar\omega$ dependence. This minimum is seen from Fig. \ref{hwdep}
to ascend with increasing model space.
Based on our results in lighter systems with larger 
spaces that show uniform convergence of this minimum, the
minimum obtained in the 
$6\hbar\omega$ model space 
is a suggested lower bound (LB) for the ground state energy. The
difference between these upper and lower bounds 
is our estimate for the `error bars' of our predictions. These error bars 
suggest reasonable convergence is attained but this requires 
verification in larger basis spaces.

\begin{table}
\extrarowheight=3pt
\tabcolsep=2pt
\caption{Binding energies (in MeV) of nuclei obtained with bare JISP16 and
   effective interaction  generated by JISP16. }
\label{bind16}
\begin{tabular}{@{}cccccc@{}}\hline
Nucleus &Nature &Bare & Effective &\parbox{9mm}{\centering$\hbar\omega$ (MeV)}
                           & \parbox{10mm}{\centering Model space} \\ \hline

$^3$H & 8.482  &8.354 & 8.496(20) &7  &$14\hbar\omega$ \\
$^3$He &7.718  &7.648 & 7.797(17) &7  &$14\hbar\omega$   \\
$^4$He &28.296 &28.297 &28.374(57) &10  &$14\hbar\omega$   \\
$^6$He &29.269  & &28.32(28) &17.5 &$12\hbar\omega$    \\
$^6$Li & 31.995 & &31.00(31) &17.5 &$12\hbar\omega$   \\
$^7$Li &39.245 & &37.59(30) &17.5 &$10\hbar\omega$ \\
$^7$Be &37.600 & &35.91(29)  &17 &$10\hbar\omega$ \\
$^8$Be &56.500 &           &53.40(10) &15 &$8\hbar\omega$ \\
$^9$Be & 58.165 & 53.54 &54.63(26) &16 &$8\hbar\omega$ \\
$^9$B  & 56.314 &51.31 &52.53(20) &16 &$8\hbar\omega$ \\
$^{10}$Be &64.977 &60.55 &61.39(20) &19 &$8\hbar\omega$ \\
$^{10}$B &64.751 &60.39 &60.95(20) &20 &$8\hbar\omega$ \\
$^{10}$C &60.321 &55.26 &56.36(67) &17 &$8\hbar\omega$ \\
$^{11}$B &76.205 &69.2 &73.0(31) &17 &$6\hbar\omega$ \\
$^{11}$C &73.440 &66.1 &70.1(32) &17 &$6\hbar\omega$ \\
$^{12}$B &79.575 &71.2 &75.9(48) &15 &$6\hbar\omega$ \\
$^{12}$C &92.162 &87.4 &91.0(49) &17.5 &$6\hbar\omega$ \\
$^{12}$N &74.041 &64.5 &70.2(48) &15 &$6\hbar\omega$ \\
$^{13}$B &84.453 &73.5 &82.1(67) &15 &$6\hbar\omega$ \\
$^{13}$C &97.108 &93.2 &96.4(59) &19 &$6\hbar\omega$ \\
$^{13}$N &94.105 &89.7 &93.1(62) &18 &$6\hbar\omega$ \\
$^{13}$O &75.558 &63.0 &72.9(62) &14 &$6\hbar\omega$ \\
$^{14}$C &105.285 &101.5 &106.0(93) &17.5 &$6\hbar\omega$ \\
$^{14}$N &104.659 &103.8 &106.8(77) &20 &$6\hbar\omega$ \\
$^{14}$O &98.733 &93.7 &99.1(92) &16&$6\hbar\omega$ \\
$^{15}$N &115.492 &114.4 &119.5(126)  &16 &$6\hbar\omega$ \\
$^{15}$O &111.956 &110.1 &115.8(126) &16 &$6\hbar\omega$ \\
$^{16}$O &127.619 &126.2 &133.8(158) &15 &$6\hbar\omega$ \\ \hline
\end{tabular}
\end{table}

Similar trends are found for most of the $p$ shell nuclei.  We
present in Table \ref{bind16} their binding energies
obtained with both bare and effective interactions. We also
quote the $\hbar\omega$ values providing the minimum 
with the effective interaction. 
The difference between the given result and the result obtained with
the same $\hbar\omega$ in the
next smaller model space is  presented in parenthesis
to  give an estimate of the convergence of our calculations. 
We quote our differences in significant figures from the 
rightmost figure of the 
stated result, omitting decimal points to save space.
The ground state energy of $A=6$, 7 and 8 nuclei converges 
uniformly from above with both the bare and effective interactions.
We present in  Table~\ref{bind16} only  the  effective interaction
results for these nuclei due to their superior convergence features. 
For these nuclei, an extrapolation based on the
fit by a constant plus exponential function 
for different $\hbar\omega$ values may be useful. For $^6$Li, this
extrapolation results in a binding energy of
31.70(17) MeV where the value in parenthesis is the uncertainty of the fit.
A similar extrapolation for $^6$He results in a binding energy of
28.89(17) MeV which is bound with respect to the $\alpha+n+n$ threshold.
We note that the bare interaction results for $A=6$ nuclei are very
close to the effective interaction ones demonstrating a remarkable
softness of the JISP16 interaction: the $^6$Li and $^6$He binding
energies are  30.94(44)
and 28.23(41) MeV respectively, the extrapolations of the bare
interaction bindings produce 
31.33(12) MeV for $^6$Li and 28.61(12) MeV for $^6$He. 
\begin{table*}
\extrarowheight=3pt
\tabcolsep=4pt
\caption{Ground state energy $E_{gs}$ and  excitation energies
$E_x$ (in MeV),  ground state point-proton rms radius $r_p$ (in  fm)
and quadrupole moment $Q$ (in $e\cdot{\rm fm^2}$) of the $^6$Li nucleus;
$\hbar\omega=17.5$ MeV.} 
\label{6LiSp}
\begin{tabular}{@{}ccccccc@{}}\hline
Interaction &\raisebox{-2ex}[0pt][0pt]{Nature} 
                         &JISP6 &JISP16 & AV8'+TM' &\!AV18+UIX &AV18+IL2 \\
Method& &NCSM, $10\hbar\omega$ \cite{JISP6} 
              &NCSM, $12\hbar\omega$ &NCSM, $6\hbar\omega$ \cite{NaO}
                &GFMC \cite{GFMC,IL2exc} &GFMC \cite{Illinois,IL2exc}\\ \hline
$E_{gs}(1^+_1,0)$ &$-31.995$ &$-31.48$  &$-31.00$ &$-31.04$
                                       &$-31.25(8)$  &$-32.0(1)$\\
$r_p$       & 2.32(3)  & 2.083
                           &2.151 &2.054 &2.46(2) &2.39(1)\\
$Q$          & $-0.082(2)$ &$\!-0.194
                        \!$  &$\!-0.0646$  &$-0.025$ &$-0.33(18)$ &$-0.32(6)$\\
$E_x(3^+,0)$ &2.186  &2.102
                           &2.529 &2.471 
                                         & 2.8(1) &2.2  \\
$E_x(0^+,1)$ &3.563  &3.348
                           & 3.701 &3.886 &3.94(23) &3.4 \\
$E_x(2^+,0)$ &4.312  &4.642
                             &5.001 &5.010 
                                           & 4.0(1) &4.2 \\
$E_x(2^+,1)$ &5.366  &5.820
                                & 6.266 &6.482   & &5.5 \\
$E_x(1^+_2,0)$ &5.65 
                        &6.86
                                   &6.573 &7.621 & 5.1(1) &5.6\\ \hline
\end{tabular}
\end{table*}

The nuclear Hamiltonian based on the {\em ab exitu} realistic
$NN$ interaction JISP16, is seen to reproduce well the binding energies of
nuclei with $A\leq 16$. The lowest state of natural parity has the
correct total angular momentum in each nucleus studied.
The experimental binding energies of all  
nuclei presented in  Table \ref{bind16}  either lie within error bars of our
predictions or are close to our suggested LB based on the effective
interaction calculations. Generally JISP16 slightly underbinds only nuclei  
in the middle of the $p$ shell. 
The difference between UB and LB is small, suggesting that JISP16
provides good convergence. However, our
error bars increase as binding energy decreases 
in a chain of isobars (cf the results for $^{13}$O and $^{13}$N).

We present in Tables \ref{6LiSp}
and  \ref{10BSp} spectra and ground state properties
of $^6$Li and $^{10}$B which 
are known \cite{NaO,GFMC,IL2exc,QMC-A910} to be
sensitive to an explicit $NNN$ interaction.
Here, the {\em ab exitu} JISP16 $NN$ interaction alone 
provides a  good description. 
The JISP16 $^6$Li  spectrum seems to be less favorable
than that provided by our JISP6 interaction 
specifically fitted to the $^6$Li  spectrum. 
However, the JISP16 $^6$Li spectrum is competitive with those of
realistic $NN+NNN$ potential models. Also, we obtain a good
description of the $^6$Li quadrupole moment $Q$ that is a recognized challenge
due to a delicate cancellation between deuteron
quadrupole moment and the $d$ wave component of the $\alpha$--$d$
relative wave function. 
We observe that $Q$ and the point-proton
rms radius $r_p$ have
a more prominent $\hbar\omega$ dependence than the binding
energy. 

\begin{table*}
\extrarowheight=3pt
\caption{Same as in Table \ref{6LiSp} but for the $^{10}$B
nucleus; $\hbar\omega=15$ MeV.}
\label{10BSp}
\begin{tabular}{ccccc}\hline
Interaction &\raisebox{-2ex}[0pt][0pt]{Nature} &JISP16 & AV8'+TM' 
                                    &AV18+IL2 \\
Method & &NCSM, 
                $8\hbar\omega$  &NCSM, $4\hbar\omega$ \cite{NaO}
                               &GFMC \cite{QMC-A910} \\ \hline
$E_{gs}(3^+_1,0)$ &$-64.751$ & 
                               $-60.14$ &$-60.57$     &$-65.6(5)$\\
$r_p$       & 2.30(12)  & 
                            2.168  & 2.168 
                                         & 2.33(1)\\
$Q$    & $+8.472(56)$ & 
                          6.484 & $+5.682$ 
                                         & $+9.5(2)$ \\
$E_x(1_1^+,0)$ & 0.718  & 
                           0.555  &0.340 
                                        & 0.9 \\
$E_x(0^+,1)$ &1.740  & 
                      1.202  &1.259 
                                         \\
$E_x(1_2^+,0)$ &2.154  & 
                         2.379 &1.216 
                                         & \\
$E_x(2_1^+,0)$ &3.587  & 
                          3.721 &2.775   
                                        &3.9 \\
$E_x(3^+_2,0)$ &4.774 & 
                        6.162   &5.971 
                                         & \\
$E_x(2_1^+,1)$ &5.164 & 
                          5.049  &5.182    
                                        & \\
$E_x(2_2^+,0)$ &5.92 & 
                       5.548   &3.987     
                                         &\\
$E_x(4^+,0)$ &6.025 & 
                       5.775  &5.229     
                                        &5.6 \\
$E_x(2_2^+,1)$ &7.478 & 
                         7.776  &7.491     \\ \hline 
\end{tabular}
\end{table*}

The  $^{10}$B properties are
also seen to be well-described with  the JISP16
interaction contrary to previous results from pure
realistic $NN$ interactions \cite{NaO,QMC-A910}. 
We note that the $^{10}$B spectrum depends 
on $\hbar\omega$ at $N_{max}=8$ but not so strongly as to alter our main
conclusions.
For example, the minimum of the  $^{10}$B ground state
corresponds to $\hbar\omega=20$ MeV while the 
minimum in the first excited state
energy  occurs at $\hbar\omega=15$ MeV. 
We  present in Table \ref{10BSp} the
$^{10}$B  properties obtained with $\hbar\omega=15$ MeV, i.e. with the
$\hbar\omega$ value corresponding to the minimum of the first excited
state since it has a more pronounced $\hbar\omega$ dependence 
than the ground state.
The $^{10}$B ground state spin was not previously reproduced
with  a pure realistic  $NN$ interaction. We observe that
our description of the  $^{10}$B spectrum is 
somewhat better than the one obtained with
the Argonne AV8' $NN$ potential and Tucson--Melbourne
TM' $NNN$ force. 
In particular,  we reproduce the ordering of
$^{10}$B levels except for the $(3_2^+,0)$ state. We
note that the $(3_2^+,0)$ state is also  
too high with the AV8'~+~TM' interaction.

In constructing ISTP \cite{ISTP}, JISP6 \cite{JISP6} 
and JISP16 potentials we adopted 
only the accepted symmetries of the $NN$ interaction and neglected 
explicit constraints such as the long-range behavior from 
meson-exchange theory.
However, this does
not mean that the JISP16 $NN$ interaction is inconsistent with
meson-theoretical forms of the $NN$ interaction. On the contrary, it is 
well-known that the one-pion exchange (OPE) dominates the $NN$ 
interaction in higher partial waves 
and the long-range behavior of $NN$ interaction
in lower partial waves. In this context, we showed in Ref. \cite{ISTP}
that our scattering wave functions in higher partial waves are nearly
indistinguishable from those of the Nijmegen-II OPE
potential. Also, in lower partial waves, our wave functions are very close
to those of  Nijmegen-II at large distances and a small difference 
is seen only at higher energies. 
Finally, we introduced the PETs of JISP6 and JISP16 only 
in lower partial waves and only in a 
few lowest oscillator components of the potential 
with a large value of $\hbar\omega=40$ MeV.
As a result, PETs reshape the wave functions at
short distances ($\lesssim 1$~fm).  Thus, the JISP16 interaction
appears to be consistent with the well-established OPE tail as embodied in the
Nijmegen-II $NN$ interaction. 

We propose our {\it ab exitu} JISP16 as a realistic  $NN$ interaction
since it describes the two-body observables with 
high precision. In addition, it provides a reasonable and economic 
description of properties of many-body nuclear systems 
in the microscopic NCSM approach.  Economy arises from
the softness of the interaction represented in a separable oscillator form.
Short distance phase-equivalent transformations adjust
the off-shell properties successfully to reduce the roles of multi-nucleon
interactions. The particular mechanism of this reduction is not clear 
at the present time.  However, our results as well as the success
of the approach of Ref. \cite{Dole}, clearly demonstrate that such a
mechanism exists and should be studied in detail. We plan to study this
with explicit $NNN$ interactions.

We conclude that the many-body nuclear
Hamiltonian obtained in our {\it ab exitu} approach is realistic from
the point of view of providing a good description 
of a wide range of nuclear data. The suggested
JISP16 $NN$ interaction 
opens  a path for extending realistic microscopic
theory to heavier nuclei, to achieve better
convergence and to obtain improved 
agreement with experiment.

This work was supported in part
by the Russian Foundation of Basic Research  grant No 05-02-17429, by 
US~NSF grant No~PHY-007-1027 
and under the auspices of the US Department of
Energy by the University of California, Lawrence Livermore 
National Laboratory under contract No.  W-7405-Eng-48 and under US~DOE grants
\mbox{DE-FG-02~87ER40371} and DE--AC02--76SF00515.

\end{document}